\documentclass[12pt]{article}

\usepackage[english]{babel}
\usepackage{amsmath,amssymb}
\usepackage{graphicx}

\numberwithin{equation}{section}

\title{Extending the nonsingular hyperscaling violating spacetimes}
\author{Yang Lei and Simon F. Ross}

\begin{document}
\maketitle

\begin{abstract}
Lifshitz and hyperscaling violating geometries, which provide a holographic description of non-relativistic field theories, generically have a singularity in the infrared region of the geometry, where tidal forces for freely falling observers diverge, but there is a special class of hyperscaling violating geometries where this tidal force divergence does not occur. We explicitly construct a smooth extension of the spacetime in this case, and explore the structure of the spacetime. We find that the spacetime has two disconnected boundaries, as in AdS$_2$. We also consider the behaviour of finite-energy excitations of the spacetime at the horizon, arguing that they will have some divergence there.
\end{abstract}

\section{Introduction}

The application of holography to the study of field theories of relevance to condensed matter systems has been a subject of intense activity in recent years (see e.g. \cite{Sean, Mcgreevy} for reviews). In particular, the application to non-relativistic theories represents an interesting extension of the usual holographic dictionary. The simplest example of this type is the Lifshitz spacetime \cite{Kachru}. The geometry is
\begin{equation} \label{lifm}
ds^2=L^2(-r^{2z}dt^{2}+r^2d\vec{x}^2+\frac{dr^2}{r^2}),
\end{equation}
where there are $d_s$ spatial coordinates $\vec{x}$. The spacetime has an isometry 
\begin{equation} \label{lifs}
t \to \lambda^z t, \quad x_i \to \lambda x_i, \quad r \to \frac{r}{\lambda}
\end{equation}
where $ z $ is called the dynamical critical exponent, which realises geometrically the anisotropic scaling symmetry found at fixed points in some condensed matter models. The case $z=1$ gives the familiar AdS spacetime, while $z \to \infty$ gives an AdS$_2 \times \mathbb{R}^{d_s}$ spacetime. These two limiting cases have a smooth extension through the apparent singularity at $r=0$ in the geometry \eqref{lifm}.  However, this is not the case in the Lifshitz spacetime, as was already noted in \cite{Kachru}, and was later stressed in \cite{Mann1,Horowitz}.  Scalar curvature invariants constructed from \eqref{lifm} are necessarily finite --- indeed, constant --- as a consequence of the Lifshitz symmetry, but there are divergent tidal forces as we approach $r=0$ along geodesic congruences. The consequences of this singularity for observers in the spacetime were explored in \cite{Horowitz}, who argued that observers near the singularity would experience large effects, although in a particular model, \cite{Bao:2012yt} found that including the effects of the matter background made the effects on a test string finite. 

The significance of the singularity in the Lifshitz metrics from the point of view of the dual field theory is less clear. As in the usual AdS/CFT correspondence, the natural observables to consider in the field theory are local correlation functions, which correspond to bulk correlators with their endpoints on the boundary of the spacetime at $r= \infty$. By causality, the calculation of these correlators only involves the region of spacetime $r > 0$, so they are not directly sensitive to the singularity. Indeed, the correlators can be calculated by analytic continuation from the Euclidean version of \eqref{lifm}, which has no divergent tidal forces. In the Euclidean solution, $r=0$ is at infinite proper distance, so the Euclidean metric in these coordinates is already geodesically complete, just as in Euclidean AdS. There is no question of extension of the solution in the Euclidean solution.

We conjecture that this singularity is reflected in the field theory in the structure of the infrared divergences appearing in scattering amplitudes. Scattering amplitudes are an intrinsically Lorentzian observable, and it is well-known that in massless theories they have infrared divergences associated with the emission of soft collinear particles. In the AdS context, non-trivial initial and final states in the Poincare patch of the geometry (corresponding to scattering amplitudes in the field theory) are associated with particles/fields crossing the Poincare horizons \cite{Balasubramanian:1999ri}. Further, in the work of Alday and Maldacena on gluon scattering amplitudes \cite{Alday:2007hr} the infrared divergence was cut off by introducing an explicit brane source in the bulk spacetime away from the Poincare horizon; the infrared divergence reappears in the limit as the cutoff brane approaches the horizon. 

To understand the role of the singularity better, it's useful to understand what is special about cases where these tidal divergences don't arise. This is relatively easy to understand in the relativistic case; scattering amplitudes are not really a good physical observable in a relativistic conformal field theory, and one should work instead with the extension of the field theory to the Einstein static universe $\mathbb R \times S^{d_s}$. The extension of the spacetime beyond the Poincare horizon seems necessarily connected to this extension of the field theory. This is also connected to the existence of special conformal transformations,  as it is the special conformal transformations that map the conformal boundary of Minkowski space to finite position (the inversion symmetry exchanges null infinity with the light cone of the origin).  

There are two non-relativistic examples where the tidal divergences also don't arise. The first is the Schr\"odinger spacetimes, which we discuss in section \ref{schr}, reviewing the extension constructed by \cite{Blau}. Schr\"odinger with $z=2$ follows the same pattern as the relativistic case: there is a special conformal symmetry, and the smooth extension of the spacetime is associated with an extension of the boundary geometry. Indeed, the bulk coordinate transformation was obtained in \cite{Blau} by  using the special conformal symmetry. However, there is a smooth extension for $z \geq 2$, and not just in the case $z=2$ where the special conformal transformation exists. This thus seems to provide an example of a solution with an extension both in the field theory and in spacetime, but without a special conformal transformation. Our new contribution to the consideration of this case is just to note that (except for the case with three bulk dimensions) these extensions are not present once we consider asymptotically Schr\"odinger spacetimes with non-zero particle number. Thus, the unexpected extensions appearing in the $z >2$ cases appear to be some special property of the field theory in an ``empty box'', where we consider a system with Schr\"odinger symmetry, but with no  particles present.

The more interesting and surprising non-relativistic example is the case  found in \cite{Edgar}, who showed that there is a particular class of hyperscaling violating spacetimes which have no tidal singularity on the horizon. This case is the main focus of our work. We will show in section \ref{ext} that these solutions have a smooth extension through $r=0$, by explicitly constructing a good coordinate system there. The dual field theory has no special conformal symmetry; indeed it doesn't even have a scaling symmetry. Furthermore, we will argue that the boundary of the extended spacetime has two disconnected components, as in AdS$_2$. Thus, the extension of the spacetime is not connected to an extension of the field theory to a larger background. 

Applying the usual holographic correspondence, we would expect such a spacetime to be dual to two copies of the field theory, with separate Hilbert spaces associated to the two boundaries. But the horizon separating the two asymptotic regions has zero cross-sectional area, so unlike in AdS$_2$, it seems problematic to interpret this geometry as corresponding to an entangled state in the two copies of the field theory. Thus, this example poses a challenge not just to our understanding of the significance of the singularities, but also to the picture advocated for example in \cite{VanRaamsdonk:2010pw,Maldacena:2013xja} that connectedness of the spacetime is dual to entanglement in the field theory. 

In section \ref{gfn}, we consider excitations of the smooth hyperscaling violating spacetimes. In AdS$_2$, it is well-known that finite-energy excitations modify the asymptotics \cite{Maldacena:1998uz}, so it is not actually possible to propagate an influence from one boundary to the other without violating the asymptotic boundary conditions.\footnote{See e.g. \cite{Balasubramanian:2010ys} for a discussion of the conceptual issues in AdS$_2$.} We consider the position-space Green's function for a source on one boundary, and while we are not able to fully calculate its form, we will argue that it has a divergence at the horizon. It is not clear whether this divergence is sufficiently strong to obstruct the extension, but it suggests there may be no physical meaning to considering correlations between the two boundaries.

While this provides a possible resolution of the puzzle, it still seems surprising even at the level of vacuum states that we can have a connection in the spacetime between the two asymptotic boundaries without any entanglement in the field theory state. It would be interesting to understand the field theory interpretation of these cases better. 

\section{Extension of Schr\"odinger spacetimes}
\label{schr}

In this section we review known results on extension of the Schr\"odinger spacetimes. This will provide a useful warmup for our later consideration of the hyperscaling violating spacetime,  and this is also an interesting example worth including in the discussion in its own right.

These spacetimes were introduced in \cite{Son:2008ye,Balasubramanian:2008dm} as duals to non-relativistic theories where the anisotropic scaling symmetry is supplemented by invariance under Galilean boosts. In the special case $z=2$, the symmetries also include a special conformal transformation. It was shown in \cite{Blau} that the Schr\"odinger solutions have a smooth extension through $r=0$ for $z \geq 2$. The extension for $z=2$ is consonant with our expectations, and indeed the smooth coordinates of \cite{Blau} were constructed by making use of the special conformal transformation. The fact that the extension continues to exist for $z >2$ is surprising, however. We will review the construction of \cite{Blau} and comment on what happens when we consider non-vacuum states in the field theory.
 
The Schr\"odinger geometry is \cite{Son:2008ye,Balasubramanian:2008dm}
\begin{equation} \label{schrm}
ds^2 = - r^{2z} dt^2 + r^2 (-2 dt d\xi + d\vec{x}^2) + \frac{dr^2}{r^2}.
\end{equation}
It represents a holographic dual of the ground state for a field theory in $d_s$ spatial dimensions with a Schr\"odinger symmetry, which includes an anisotropic scaling symmetry and Galilean boosts. Realising this extended symmetry requires an extra dimension. In particular the addition of the $\xi$ coordinate enables us to realize the conserved particle number appearing in the Schr\"odinger algebra as momentum in the $\xi$ direction. As in the Lifshitz spacetime, this solution has an apparent singularity at $r= 0$. An extension of the spacetime beyond $r=0$ was found in \cite{Blau} for $z \geq2$. For $z=2$, their construction was based on the special conformal symmetry $C$ which appears for this choice of dynamical exponent. They define a new timelike coordinate $T$ such that $\partial_T  = H + C = \partial_t + C$. This led them to define the new coordinates $(T,R,\vec{X},V)$ given by
\begin{equation} \label{bcoord}
t = \tan T, \quad r = \frac{\cos T}{R}, \quad \vec{x} = \frac{\vec{X}}{\cos T}, 
\end{equation}
\begin{equation} \label{bc2}
\xi = V + \frac{1}{2} (R^2 + \vec{X}^2) \tan T.
\end{equation}
In these new coordinates, the metric for $z=2$ is 
\begin{equation} \label{blaum}
ds^2 = - \frac{dT^2}{R^4} + \frac{1}{R^2} (-2 dT dV - (R^2 + \vec{X}^2) dT^2+ dR^2 + d\vec{X}^2).
\end{equation}
The null surfaces at $r=0$ in the original metric correspond to surfaces $\cos T =0$ which are evidently smooth in the new coordinates. There is still an apparent singularity at $R \to \infty$ in the new coordinates, but because of the harmonic potential in $g_{TT}$, geodesics are prevented from reaching $R \to \infty$, so this new spacetime is actually geodesically complete. From the point of view of the boundary at $r = \infty$ ($R=0$),  the extension adds regions to the future and past of the existing boundary. For $z=2$, this extension of the boundary can be understood as a result of the special conformal transformation. Thus, the case with $z=2$ has the same qualitative structure as for AdS. 

The surprise is that this coordinate transformation also provides a smooth extension of the spacetime for $z >2$. Applying the same coordinate transformation in the case of general $z$ gives 
\begin{equation}
ds^2 = - (\cos T)^{2z-4}  \frac{dT^2}{R^4} + \frac{1}{R^2} (-2 dT dV - (R^2 + \vec{X}^2) dT^2+ dR^2 + d\vec{X}^2).
\end{equation}
Thus, for $z >2$, the extension is still smooth at $\cos T = 0$. The geometry no longer has a  $T$-translation symmetry, which is a consequence of the absence of the special conformal transformation, but there is no obstruction to the extension, and the picture from the point of view of the causal boundary is the same as before. This is an example where the field theory unexpectedly has a smooth extension; there was no symmetry in the theory in the $t, x, \xi$ coordinates which suggested that it would be reasonable to treat $t \to \infty$ as being at finite position, but the coordinate transformation \eqref{bcoord}, which maps this to finite $T$, gives a smooth bulk geometry with an extended boundary.

A natural suspicion is that this smooth extension is a special feature of the vacuum solution. To say that an extension really exists in the field theory, we would like to see that there are excitations of the geometry which remain smooth in global coordinates, corresponding to non-trivial states of the field theory on the extended spacetime. In the next two subsections we consider two kinds of excitations; changes in the state in a sector with a given particle number, and changes in the conserved particle number of the field theory. 

\subsection{Excitations: mode solutions}

We want to consider excitations about the Schr\"odinger solution, and look for excitations which remain smooth in global coordinates. In this section we consider normalizable mode solutions, corresponding to excited states of the field theory, following \cite{Blau}. In appendix \ref{schrg}, we give some new results on position space Green's functions in this spacetime. 

The simplest thing to do is to consider mode solutions in the original coordinates. However, unlike on a black hole spacetime, there are no mode solutions which are regular at the horizon; that is, there is no analogue of ingoing modes on the Schr\"odinger background. This is trivial to see. The mode solutions in the original coordinates are 
\begin{equation} \label{oldm}
\phi = e^{-i m \xi + i \omega t + i \vec{k} \cdot \vec{x}} f(r).
\end{equation}
As $r \to 0$ along a generic ingoing geodesic, all of $t, \xi, \vec{x}$ diverge. For example, if we take $r \to 0$ keeping $V$ finite, $\xi$ will blow up. Whatever the dependence of $f(r)$ on $r$ is, the assumption that the mode depends separately on $t, \vec{x}$ and $r$ means that the dependence on $\xi$ cannot become a dependence on the finite coordinate $V$.\footnote{In a black hole spacetime, the mode solutions are $\phi = e^{i \omega t} f(r)$, and the divergence of $t$ on the future horizon can be cancelled by choosing an ingoing solution for $f(r)$, so that $\phi \approx e^{i \omega u}$ near the horizon, where $u$ is an ingoing Eddington-Finkelstein coordinate. The point of this comment is that because more coordinates blow up on the Schr\"odinger horizon, no such cancellation can be engineered just by choosing $f(r)$.} Thus there are no mode solutions in the original coordinates that are regular at the horizon. This is of course no obstruction to the existence of smooth solutions; it just says that the modes \eqref{oldm} are not a good basis for constructing them.  

For the Schr\"odinger solution with $z=2$, the geometry has enough symmetry to allow us to solve for mode solutions in the new coordinates. This analysis was carried out in \cite{Blau}, for the solutions for a probe scalar field on the Schr\"odinger background in the new coordinates. If we solve the massive Klein-Gordon equation with mass $\mu$ in the new coordinate system, the solutions can be decomposed in modes as 
\begin{equation} \label{newm}
\phi = e^{-iET} e^{-imV} Y_{L}(\theta_i) \varphi_{L,n}(\rho) \phi_{L,n}(R),
\end{equation}
where $\rho, \theta_i$ are spherical polar coordinates on the spatial $\vec{X}$ coordinates, $Y_L(\theta_i)$ are the appropriate spherical harmonics, and $\varphi_{L,n}(\rho)$ is given in terms of a generalized Laguerre polynomial. The radial function $\phi_{L,n}(R)$ satisfies 
\begin{equation}
\phi'' - \frac{d_s+1}{R} \phi' + (2 Em - 4m (n+\frac{L}{2}+\frac{d_s}{4}) - m^2 R^2 - \frac{(m^2 + \mu^2)}{R^2} )\phi =0. 
\end{equation}
The solutions of this equation can be written in terms of confluent hypergeometric functions. The two independent solutions near $R \to \infty$ are 
\begin{equation}
\phi \sim e^{\pm \frac{1}{2} m R^2}. 
\end{equation}
Following \cite{Blau}, the boundary condition is taken to be that we keep only the exponentially damped falloff in the limit $R \to \infty$. The regular solution is then 
\begin{equation}
\phi = e^{-\frac{1}{2} m R^2} R^{\Delta_+} U(a, b, mR^2), 
\end{equation}
where $U(a,b,mR^2)$ is Tricomi's confluent hypergeometric function, and 
\begin{equation}
a = \frac{1}{2} (1+\nu) + n + \frac{L}{2} + \frac{d_s}{4} - \frac{E}{2} , \quad b= 1+\nu. 
\end{equation}

This solution is clearly regular in the interior of the spacetime. However, it only has an interpretation as a change in the state of the field theory if it only excites the normalizable (fast fall-off) part of the field near the boundary. The Tricomi function is 
\begin{equation}
U(a,b,z) = \frac{\Gamma(1-b)}{\Gamma(a-b+1)} {}_1 F_1(a,b,z) + \frac{\Gamma(b-1)}{\Gamma(a)} z^{1-b} {}_1 F_1 (a-b+1,2-b,z). 
\end{equation} 
The regular solution is purely normalizable at infinity if the first term is absent, which can happen if we encounter a pole in the $\Gamma$ function in the demoninator, that is if $a-b+1$ is a negative integer.  This will select a discrete set of energies $E$ (as found in \cite{Blau:2010fh}),
\begin{equation}
E = 1+\nu + 2n + L + \frac{d_s}{2} + r, \quad r \in \mathbb{N}.
\end{equation}
Thus the situation in $z=2$ Schr\"odinger is very similar to that in AdS. There is a discrete spectrum of smooth mode solutions with respect to the new coordinates, and we can describe smooth excitations above the vacuum state, at least at linear order in perturbations, by considering linear combinations of these modes. 

It is difficult to extend this analysis to $z >2$, as the geometry now has no time-translation symmetry in $T$, so we cannot Fourier transform in the $T$ direction. Solving the wave equation in the new coordinates would therefore requiring solving a PDE. It would be interesting if this problem could be shown not to have smooth solutions, as this would indicate a difference between the $z=2$ and $z >2$ cases. We will not pursue this further as we will see in the next section that both $z=2$ and $z >2$ encounter a problem when we consider non-zero particle number.

\subsection{Excitations: nonzero particle number} 

Since the Schr\"odinger algebra contains a conserved particle number, in addition to asking if the extension through $r=0$ applies to excitations above the ground state, it is also natural to ask if it applies to the ground state in sectors of the theory with non-zero values of the particle number. Here we will consider what happens for uniform distributions of particle number, as one would expect in the ground state in a sector of fixed particle number. 

For $z=2$, one might think that exciting non-zero particle number would allow us to preserve the smoothness at $r=0$, since the particle number operator $N$ is central in the algebra, so it commutes with both the dilatation and the special conformal transformation. However, from the geometric point of view the relevant quantity is not the total particle number but the local particle number density $\rho$; it is the dimension of this local operator that will determine the effect of particle number on the bulk spacetime. For $z=2$, the particle number density $\rho$  has dimension $d_s$, so we would expect that giving it an expectation value will produce a deformation of the spacetime whose effect is more pronounced in the IR, modifying the structure of \eqref{schrm} at $r=0$. 

This is indeed what we find if we consider the geometries obtained by taking the zero-temperature limit of the black hole solutions for $d_s=2$ found in \cite{Herzog:2008wg,Adams:2008wt,Maldacena:2008wh} while holding the particle number fixed. The limiting geometry (in string frame) is 
\begin{equation}
ds^2 = k(r)^{-1} (-r^4 dt^2 + \frac{\gamma^2}{r^2} d\xi^2 -2 r^2 dt d\xi) +  ( r^2 d\vec{x}^2 + \frac{dr^2}{r^2} ), 
\end{equation}
where $k(r) = 1 + \frac{\gamma^2}{r^2}$. The spacetime is asymptotically Schr\"odinger, with the $1/r^2$ falloff for the deviations expected for a non-zero particle number density. We can see that the introduction of the non-zero density indeed deforms the spacetime in the IR; this solution is now singular at $r=0$. This is again a tidal divergence, with Riemann tensor components like $R_{0i0i}$ diverging in a parallelly propagated orthonormal frame along ingoing geodesics: \begin{equation}
R_{0i0i}=\frac{2\gamma^2E^2}{r^6}+(1+P_\xi^2).
\end{equation}
This component is finite if the density $\gamma$ vanishes while becomes divergent in the finite density spacetime. Thus, there is no smooth extension through $r=0$ for these solutions with non-zero particle number. We should note that there is also a divergent tidal force if we consider the metric in Einstein frame. Although we do not have explicit solutions for $z >2$, we would expect a similar logic to apply there as well. 

The exception to the preceding discussion is Schr\"odinger spacetimes with three bulk dimensions, as then $d_s=0$, and non-zero particle number produces a marginal deformation of the geometry. Indeed, in this case the Schr\"odinger solution has been identified with the null warped AdS$_3$ geometry, and the solution with non-zero particle number is the spacelike warped AdS$_3$ geometry \cite{Anninos:2008fx}, with metric
\begin{equation}
ds^2 = -r^4 dt^2 -2 r^2 dt d\xi + \gamma^2 d\xi^2 + \frac{dr^2}{r^2} .
\end{equation}
This metric is a fibration over AdS$_2$, as can be made manifest by defining $\rho = r^2$ and $\bar t = \frac{2\sqrt{\gamma^2 +1}}{\gamma} t$, so
\begin{equation}
ds^2 = \frac{1}{4} ( - \rho^2 d\bar t ^2 + \frac{d\rho^2}{\rho^2} ) + \gamma^2 (d\xi - \frac{\rho}{2(1+\gamma^2)} d\bar t)^2. 
\end{equation}
Here, the singularity at $r=0$ can be resolved by passing to global coordinates for the AdS$_2$ factor, both for vanishing and for non-vanishing particle number. Thus, the extension of the spacetime exists for non-zero particle number. On the other hand, the fact that the geometry involves AdS$_2$ implies that the excitations in a sector of given particle number considered in the previous section fail once we take into account back-reaction (at zero or non-zero particle number), since AdS$_2$ does not have finite excitations which are asymptotically AdS$_2$ on both asymptotic boundaries in the global coordinates \cite{Maldacena:1998uz}.

Another caveat to the argument is that it applies to solutions with finite particle number density; it may be that there could be some solutions with finite total particle number (in a spatially infinite field theory) which remain smooth at $r=0$. However, as such solutions would necessarily be time-dependent it is significantly more difficult to analyze the question, and it is the case of finite particle number density which is of real practical interest. 

\section{Hyperscaling violating spacetimes}
\label{ext}

In this section, we turn to our main subject, the non-singular hyperscaling violating spacetimes. We will first review the general class of spacetimes, and briefly discuss the non-singular case, before explicitly constructing a smooth extension for this case through the horizon at $r=0$ and discussing the resulting global structure. 

The general hyperscaling violating geometry has the form \cite{Ogawa,Liza,Dong}
\begin{equation} \label{hsvm}
ds^2=\frac{1}{\bar r^{2\theta/d_s}}(- \bar r^{2z} dt^2 + \bar r^2 d\vec{x}^2 + \frac{d \bar r^2}{\bar r^2}). 
\end{equation}
The $\theta=0$ case is the Lifshitz spacetime \eqref{lifm}. For $\theta \neq 0$, the isometry under \eqref{lifs} is broken; the metric has an overall scaling under this transformation, $ds^2 \to \lambda^{2\theta/d_s} ds^2$. Theories with such a hyperscaling violation have a characteristic thermodynamic behaviour which is that of a theory living in $d_s - \theta$ dimensions. As a result, it has been suggested that these metrics with $\theta = d_s-1$ have a thermodynamic structure which  may be a useful model for a field theory with a Fermi surface \cite{Liza} (as the effectively one-dimensional behaviour  reproduces the behaviour near a Fermi surface). 

In what follows, we will work in a coordinate system where we define the radial coordinate as the proper size of the spatial directions; the metric is then (up to an overall scale)
\begin{equation} \label{hsvm2}
ds^2= - r^{2 m} dt^2 + \frac {d r^2}{r^{2n}} + r^2 d\vec{x}^2. 
\end{equation}
The relation between the coordinates is $r \sim \bar r^{2(d_s-\theta)/d_s}$, and 
\begin{equation}
m = \frac{d_s z-\theta}{d_s - \theta}, \quad n = \frac{d_s}{d_s - \theta}. 
\end{equation} 
As for the Lifshitz solution, these geometries generically have a curvature singularity at $r=0$, but there is an exception; as noted in \cite{Edgar}, the case $m= n \geq 2$ has no diverging tidal forces as we approach $r=0$. The curvature components in a parallelly propagated frame include \cite{Mann2}
\begin{equation}
R_{0i0i}=\frac{r^{2n-2m-2}[2(m-n)E^2+r^{2m-2}[(2n-2)p^2+2nr^2]]}{2}
\end{equation}
\begin{equation}
R_{1i1i}=\frac{r^{2n-2m}[2(m-n)E^2-2mr^{2m-2}(p^2+r^2)]}{2(p^2+r^2)}
\end{equation}
\begin{equation}
R_{0i1i}=\frac{(2m-2n)Er^{2n-2m-1}}{2(p^2+r^2)}\sqrt{E^2-r^{2m}(1+\frac{p^2}{r^2})}
\end{equation}
In general, the parameters are restricted to $m \geq n$ by the null energy condition (generalizing the familiar restriction to $z \geq 1$ in the Lifshitz case). Given this, we can see that for these components of the Riemann tensor to remain regular as $r \to 0$, we must have $m = n \geq 2$. It can be checked that given this condition, all the components of the Riemann tensor remain finite in the limit \cite{Edgar,Mann2}. 

The non-divergent case is special in the sense that it saturates the bound from the null energy condition.\footnote{Although the null energy condition is satisfied, this spacetime does require negative energy densities, and as noted in  \cite{Mann2} it is not straightforward to construct reasonable matter Lagrangians that give rise to it as a solution. Since our interest is mainly in using this example to test our general understanding, rather than to advance it as a physically interesting model, we have not attempted to address this issue.}  For two spatial dimensions, the choice $z= 3/2$, $\theta = 1$, which gives $m = n = 2$, was also previously identified as special because it gives rise to a logarithmic violation of the area law for entanglement entropy \cite{Liza}, so it may be interesting for modelling Fermi liquids holographically. 

For simplicity, we will focus mainly on the case $m=n=2$, and comment briefly on the extension to larger values at the end. The metric is 
\begin{equation}
ds^2=-r^4dt^2+\frac{dr^2}{r^4}+r^2 dx_i^2.
\end{equation}
The fact that the metric is non-singular precisely when $g_{rr} = 1/g_{tt}$ suggests that it will be useful to introduce a tortoise coordinate $r_*$ such that $dr_* = g_{rr} dr$, as in the Schwarzschild spacetime. Indeed, if we define 
\begin{equation}
u=t-\frac{1}{3r^3}, 
\end{equation}
the metric becomes 
\begin{equation} \label{umet}
ds^2=-r^4du^2+2dudr+r^2dx_{i}^2.
\end{equation}
The $u,r$ part of the metric is regular, but this cannot be the end of the story, as the metric in the $x_i$ directions is still degenerating as $r \to 0$. We can get some insight into the situation by considering the behaviour of the geodesics. The conserved energy and momentum along the geodesics are $E=r^4 \dot{t}$ and $p_i=r^2\dot{x_i}$. Thus 
\begin{equation}
\dot{r}^2=E^2-p^2r^2-\epsilon r^4,
\end{equation}
where $\epsilon =1$ for timelike geodesics and $\epsilon =0$ for null geodesics,
which implies \begin{equation} \label{tdot}
\frac{dt}{dr}=\frac{\dot{t}}{\dot{r}}=-\frac{E}{r^4\sqrt{E^2-p^2r^2-\epsilon r^4}},
\end{equation}
\begin{equation} \label{xdot}
\frac{dx_i}{dr}=\frac{\dot{x_i}}{\dot{r}}=-\frac{p_i}{r^2\sqrt{E^2-p^2r^2-\epsilon r^4}},
\end{equation}
where we are considering ingoing geodesics. Near $ r=0 $,
 \begin{equation}
t\approx \frac{1}{3r^3}+\frac{1}{2}\frac{p^2}{E^2r}+ \ldots
\end{equation}
and
\begin{equation}
x_i\approx \frac{p_i}{Er}+ \ldots, 
\end{equation}
where the terms not written explicitly are bounded as $r \to 0$. For null geodesics,  we can explicitly integrate \eqref{tdot} and \eqref{xdot} to obtain 
\begin{equation}
t =  \frac{(E^2 + 2 p^2 r^2) \sqrt{E^2 - p^2 r^2}}{3 E^3 r^3} + t_0 
\end{equation}
and
\begin{equation}
x_i =  p_i \frac{\sqrt{E^2 - p^2 r^2}}{E^2 r}  + x_{i0}. 
\end{equation}
If we introduce $\bar p_i = p_i/E$, this can be rewritten as 
\begin{equation}
t=    \frac{(1 + 2 \bar p^2 r^2) \sqrt{1 - \bar p^2 r^2}}{3 r^3} + t_0, \quad x_i =  \bar p_i \frac{\sqrt{1 - \bar p^2 r^2}}{r}  + x_{i0}.  
\end{equation}

We see that the ingoing coordinate $u$ is finite (in fact constant) along the radial null geodesics with $\bar p =0$, as in Eddington-Finkelstein coordinates on a black hole. However, for the general geodesics with $p \neq 0$, the coordinate transformation has removed the leading divergence in $t$ but both $u$ and the spatial coordinates $x_i$ diverge like $r^{-1}$ near $r=0$. 

To remove these divergences, we define 
\begin{equation}
X_{i}=rx_{i}
\end{equation}
and
\begin{equation} \label{Tcoord}
T=u-\frac{X_i^2}{2r}=t-\frac{1}{3r^3}-\frac{1}{2}rx_i^2 . 
\end{equation}
The metric can be written in a simple form by introducing polar coordinates $(R,\theta_a) $ in the transverse $X_i$ space:
\begin{eqnarray} \label{smooth} 
ds^2&=&-r^4dT^2-\frac{1}{4}R^4dr^2+(1-R^2r^2)dR^2+R^2d\Omega_{d_s-1}^2 \\
&& +(2+R^2r^2)dTdr+R^3rdRdr-2Rr^3dRdT. \nonumber
\end{eqnarray}
We can see that the components of the metric remain finite at $r=0$ in these coordinates; in addition, the determinant of the metric is 
\begin{equation}
\det{g_{\mu\nu}}=-R^{2(d_s-1)},
\end{equation}
which is finite at $r=0$, so the inverse metric is also smooth there. Thus, these coordinates provide a smooth extension of the metric through $r=0$. The surface $r=0$ is a null hypersurface,  a smooth event horizon. We have constructed ingoing coordinates, allowing us to smoothly cross the future horizon at $t \to  \infty$ as $r \to 0$; we could similarly construct outgoing coordinates by taking 
\begin{equation}
T' = t + \frac{1}{3r^3} + \frac{1}{2} r x_i^2.
\end{equation}
Since the metric \eqref{smooth} is invariant under $r \to -r$, $T \to -T$, we see that the region $r <0$ is isometric to the region $r >0$. 

This method can also be generalized to other $ n \ge 2$ cases by taking 
\begin{equation} \label{nTcoord}
T=t-\frac{1}{2n-1}r^{-(2n-1)}-\frac{r}{2}x_i^2, \quad X_i = r x_i, 
\end{equation}
which gives
\begin{eqnarray}
ds^2&=&-r^{2n}dT^2+(1-r^{2(n-1)}R^2)dR^2-\frac{1}{4}r^{2(n-2)}R^4dr^2+R^2d\Omega_{d_s-1}^2\\
&& +(2+r^{2(n-1)}R^2)dTdr-2r^{2n-1}RdTdR+r^{2n-3}R^3dRdr.  \nonumber
\end{eqnarray}
Note that as expected, this provides a smooth extension only for $ n \ge 2 $.

\subsection{Global structure}

To understand the meaning of this extension of the geometry from the point of view of the dual field theory, we would like to understand the relation between the regions $r >0$ and $r <0$; in particular, we want to understand the relation between their asymptotic boundaries at large $r$, where we conventionally think of the field theories as living. (More precisely, as the hyperscaling violating  spacetime is singular as $r \to \infty$, we should introduce an explicit cutoff and work on a surface of constant $r= r_0$). In an AdS$_d$ spacetime for $d >2$, when we extend the Poincare patch to global coordinates, the boundary is connected, and there is a single Hilbert space for the dual field theory\footnote{The extension for the Schr\"odinger spacetimes reviewed above is also of this form.}; by contrast, in a black hole spacetime or in AdS$_2$, there are two disconnected boundaries, which have separate field theory Hilbert spaces associated with them. The field theory dual in those cases is some entangled state in two copies of the field theory. We would like to know whether our hyperscaling violating spacetime is of the former or of the latter type. 

In our smooth coordinates \eqref{smooth}, the spacetime certainly does not look connected, but this may be just a defect of our coordinates. To consider this question in a more coordinate-independent manner, we will consider the causal structure of the spacetime. In the cases where the boundary is connected, an initial time slice in the boundary is a Cauchy surface for the full extended boundary in the field theory, and the whole of the boundary lies either to the future or to the past of this initial time slice. So if we find that there are points on the boundary which are not in the future or past of the initial data slice in one asymptotic region of the hyperscaling violating spacetime, we can conclude that the extension of the spacetime does not correspond simply to further evolution of the field theory state defined on that initial slice, but must instead involve some extension of the field theory Hilbert space.

We are therefore interested in considering the future and past of an initial time slice, which in the bulk spacetime corresponds to a constant $t$ slice of the boundary. Thus, we want to find $I^{\pm}(r=r_0, t = t_0)$. We can see from \eqref{tdot} that motion in the $x_i$ directions restricts the motion in $r$, so the future or past of  $r=r_0$, $t=t_0$ will be bounded by the radial null geodesics. The ingoing/outgoing coordinates $u, v = t \mp \frac{1}{3r^3}$ are constant along the radial null geodesics, and the $x_i$ are constant, so $T$, although not constant, remains bounded. Thus, the ingoing radial null geodesics from $r=r_0$ will intersect the surface $r = -r_0$ beyond the future horizon at some finite value of $t$. Thus, there is indeed a part of this new asymptotic region which is acausal with respect to   the initial time surface. (No part of the region beyond the future horizon is to the past of the initial surface.) 

\begin{figure}[htbp]
\small
\centering
\includegraphics[width=5cm]{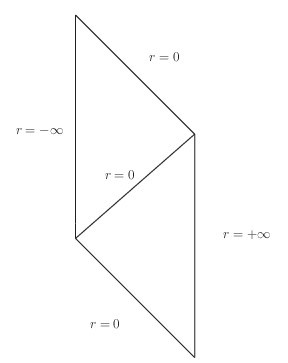}
\caption{A qualitative depiction of the causal structure in the region covered by the ingoing coordinates. Note that although we draw the $T,r$ space, the geometry does not have a translational symmetry in the transverse space in the coordinates regular at the horizon, so this is not a true Penrose diagram.}
\label{fig:diagram}
\end{figure} 

This implies that the structure of the spacetime is qualitatively similar to that of AdS$_2$, as depicted in figure \ref{fig:diagram}; there is a separate boundary at $r <0$, disconnected from the boundary at $r >0$.\footnote{It is not clear if successive boundary regions at $r >0$ are connected, as they would be in AdS$_2$; our construction has not given us a single coordinate patch covering two such regions.} 

If we follow the usual holographic dictionary, we would associate these two asymptotic boundaries with two copies of the field theory Hilbert space. Now an interesting problem is that the horizon at $r=0$ has vanishing cross-sectional area, so it is difficult to interpret the geometry as dual to an \textit{entangled} state in two copies of the field theory. If we assumed the usual Ryu-Takayanagi prescription applied, the entropy density in the reduced density matrix obtained by tracing over one of the boundaries should be given by the area of the horizon, as this is clearly an extremal surface \cite{Ryu:2006bv,Hubeny:2007xt}. The field theory coordinates are $t, x^i$, so the vanishing of $g_{x^i x^i}$ at $r=0$ in \eqref{umet} appears to say that the reduced density matrix has zero entropy density.\footnote{The horizon has a non-degenerate metric on the surfaces of constant $T$ in \eqref{smooth}, $ds^2_{r=0,T=const} = dR^2 + R^2 d\Omega_{d_s-1}^2$, but since finite $R$ at $r=0$ corresponds to infinite values of $x^i$,  this is not naturally related to the entropy density in the field theory. However, the rules for such cases with non-compact horizons are perhaps not entirely clear. We can't easily resolve the problem by compactifying the $x_i$ coordinates, as this would spoil the smoothness at the horizon, as in the Poincare patch in AdS.} Thus, the state of the field theory asociated to this spacetime would seem to have no entanglement, contrary to the general conjectures in \cite{VanRaamsdonk:2010pw,Maldacena:2013xja}.

\section{Excitations of the smooth spacetime}
\label{gfn}

The smooth extension of the spacetime indicates that the ground state of the field theory can be thought of as naturally defined on the full asymptotic boundary of the spacetime, rather than just on the boundary in the original $r >0$ region. As in the Schr\"odinger example, it is then interesting to ask if this extension has meaning also for excited states.  We will first consider looking for mode solutions of this equation in the different coordinates, and then consider a Green's function for an operator insertion on the boundary. 

\subsection{Scalar fields in the static coordinate}

In the original static coordinates, we can consider the plane wave modes
 \begin{equation}
\phi(t,r,x,y)=e^{-i\omega t+i\vec{k} \cdot \vec{x}}R(r).
\end{equation}
The Klein-Gordon equation $\nabla^2 \phi - m^2 \phi =0$ then reduces to an ODE,
\begin{equation}
\frac{1}{r^{d_s}}\partial_r(r^{4+d_s}\partial_r R)+\frac{\omega^2}{r^4}R(r)-\frac{k^2}{r^2}R(r) - m^2 R(r) =0.
\end{equation}
The near horizon region is at $r \to 0$ and boundary is at $ r \to \infty $. We can't solve this equation in closed form in general, but we are interested in the behaviour in the near horizon region, where only the first two terms are relevant, and there is an approximate solution in terms of Bessel functions, 
\begin{equation}
R(r) \approx r^{-\frac{d_s+3}{2}} (C_1 J_{\frac{3+d_s}{6}}(\frac{\omega}{3r^3}) + C_1 J_{-\frac{3+d_s}{6}}(\frac{\omega}{3r^3})).
\end{equation}
Near the horizon, the asymptotic expansion of the Bessel function is
\begin{equation}
J_\nu(x) \sim \sqrt{\frac{2}{\pi x}}\cos(x-\frac{\nu \pi}{2}-\frac{\pi}{4}),
\end{equation}
so we have the near horizon behaviour 
\begin{equation}
R(r) \sim \sqrt{\frac{2}{\pi}}r^{-\frac{d_s}{2}} e^{\pm i \omega/3r^3}
\end{equation}
as $r \to 0$. There is an overall $r^{-d_s/2}$ divergence, but leaving that aside, the $e^{i \omega/3 r^3}$ behaviour here is reminiscent of a black hole; it indicates that we could define ``ingoing" and ``outgoing" modes behaving as $e^{i \omega u}$, $e^{i \omega v}$, where $u, v = t \mp 1/3r^3$. However, while $u$ would remain finite as we approach the horizon along radial null geodesics, it would diverge as we approach the horizon along more generic directions. Thus, unlike in a black hole spacetime, and like in the Schr\"odinger example, there are no individual mode solutions which are well-behaved on the horizon. The assumption that the dependence on $t, r$ and $\vec x$ separates immediately implies that the modes cannot become functions of $T$ as we approach the horizon. 

As in the Schr\"odinger case, this tells us nothing about the smoothness of the extension, but just indicates that these modes do not provide a good basis near the horizon. 

\subsection{The scalar fields in new coordinate}

We can attempt to look for solutions of the Klein-Gordon equation in the new regular coordinates. However, this is more difficult, as there are no additional symmetries which are manifest in the new coordinates, so the wave equation does not separate in these coordinates. 

The inverse metric is  
 \begin{equation}
\left(\begin{array}{ccc}
\frac{1}{4}R^4 & 1-\frac{1}{2}r^2R^2 & -\frac{1}{2}rR^3 \\
1-\frac{1}{2}r^2R^2 & r^4 & r^3R \\
-\frac{1}{2}rR^3  &  r^3R & 1+r^2R^2
\end{array}
\right),
\end{equation}
so the equation of motion in the new coordinates is 
\begin{eqnarray}
&&\frac{1}{4}R^4\partial_T^2\phi+(2-r^2R^2)\partial_T \partial_r \phi- rR^3\partial_T\partial_R \phi+r^4\partial_r^2\phi+2r^3R\partial_R\partial_r \phi \\ &+& (1+r^2R^2)\partial_R^2 \phi-\frac{d_s+4}{2}rR^2\partial_T \phi +(d_s+4)r^3 \partial_r \phi \nonumber
 \\ &+& (\frac{d_s-1}{R}+(d_s+4)r^2R)\partial_R \phi+\frac{\partial_\Omega^2 \phi}{R^2}  -m^2\phi =0. \nonumber 
\end{eqnarray}
The $\Omega$ stands for all the angular parts which have $d_s-1$ dimensions. The $\Omega$ dependence is separable (as a consequence of the rotational symmetry in the $X_i$ plane), and we can take advantage of the time translation invariance in $T$ to Fourier transform in the $T$ direction, so we can write
 \begin{equation} 
 \phi=e^{i\alpha T}Y_{L}(\Omega)H(r,R).
 \end{equation}
then we can arrange the equation into
\begin{eqnarray} \label{generalequation}
&&r^4\frac{\partial^2H}{\partial r^2}+(1+r^2R^2)\frac{\partial^2 H}{\partial R^2} +2r^3R \frac{\partial^2H}{\partial r \partial R}+ (-i\alpha r^2R^2+(d_s+4)r^3+2i\alpha)\frac{\partial H}{\partial r} \\
&+& (-i\alpha rR^3+\frac{d_s-1}{R}+(d_s+4)r^2R)\frac{\partial H}{\partial R} \\ &+&\left( -\frac{1}{4}\alpha^2R^4-\frac{d_s+4}{2}i\alpha rR^2-\frac{L}{R^2}-m^2 \right) H=0,
\end{eqnarray}
but the $r$ and $R$ dependence in this equation does not separate, so it is not possible to make further progress analytically in general.  It is possible to separate the equation for $\alpha =0$, but this essentially reduces to the special case $\omega = 0$ of the previous analysis in the original coordinates. 

It would be interesting to investigate this equation numerically. For each spherical harmonic, one should look for values of $\alpha$ such that the solution is a regular function of $r, R$ which is purely normalizable as $r \to \pm \infty$. This seems a challenging numerical problem, so we leave it for future work and in the next section we turn to an alternative approach, studying the Green's functions for sources on the boundary. 

\subsection{Green's function}

As an alternative to the mode solution analysis, which corresponds to considering excitations of the incoming initial state at past infinity, we can consider an excitation created by acting with some localized source on the boundary. That is, we can ask if the boundary to bulk Green's function is smooth at the horizon. We will consider first a spatially uniform source, where we can explicitly find the Green's function analytically, and we can gain some understanding of its structure. We will then argue that the Green's function for a spatially localized source has a similar, albeit weaker, singularity at the horizon, although we can't do the full calculation of the Green's function explicitly in this case. It may be useful to read this section in conjunction with appendix \ref{schrg}, where the same calculation is done for $z=2$ Schr\"odinger, as in that case the calculation can be carried out explicitly in full. We do a calculation for the hyperscaling violating solution with $m=n=2$. The cases with $m=n>2$ are similar; the calculation is summarized in appendix \ref{gen}.

The hyperscaling violating spacetimes do not have a scaling symmetry; instead scaling the coordinates produces an overall rescaling of the metric. However, if we consider massless fields, this is sufficient to produce a simplification in the form of the Green's function. We will therefore restrict to the consideration of massless fields. The spacetime has a real Euclidean section defined by analytically continuing $ t \to -i \tau$, so we define the Green's function in the Lorentzian spacetime by analytic continuation from this Euclidean section.  In the Euclidean spacetime, the massless equation is 
\begin{equation}
\frac{1}{r^{d_s}}\partial_r(r^{4+d_s}\partial_r \phi)+\frac{1}{r^4} \partial_\tau^2 \phi+\frac{1}{r^2}\partial_i^2 \phi=0.
\end{equation}
This equation has a symmetry under the scaling transformation
\begin{equation} \label{lscale}
r \to \lambda^{-1}r, \qquad \tau \to \lambda^3 \tau, \qquad x_i \to \lambda^2 x_i, \qquad ds^2 \to \lambda^2 ds^2,
\end{equation}
as the scaling of the metric comes out as an overall factor in this massless equation.  
 
We consider first as a warmup a source which is smeared over the spatial directions, where we can calculate the Green's function exactly. By translation invariance in the original coordinates, we take the source to be at $\tau=0$, so that the boundary condition is 
\begin{equation}
\lim_{r\to +\infty}\phi =C \delta(\tau).
\end{equation}
The solution with this boundary condition will be independent of the $x_i$. The delta-function in the boundary conditions breaks the symmetry under the scaling \eqref{lscale}, but it transforms covariantly, so the solution should behave as $\phi(\lambda^3 \tau, \lambda^{-1} r) = \lambda^{-3} \phi(\tau, r)$. Thus, the solution should have the form $\phi(r,t)=r^3 f(r^3 \tau)$, and the problem reduces to an ODE,  
\begin{equation}
(9x^2+1)f''(x)+(36+3d_s)xf'(x)+(18+3d_s)f(x)=0,
\end{equation}
where $x=r^3 \tau$. The solution satisfying our boundary conditions is 
\begin{equation} 
f(x)=\frac{C}{(9x^2+1)^{1+\frac{d_s}{6}}}
\end{equation}
that is,
\begin{equation} 
\phi =\frac{C' r^3}{(9r^6 \tau^2+1)^{1+\frac{d_s}{6}}}.
\end{equation}
This solution satisfies the boundary conditions because it vanishes as $r \to \infty$ for $t \neq 0$, and the scaling form $\phi = r^3 f(r^3 \tau)$ automatically implies that the integral $\int \phi d\tau$ over a surface of constant $r$ is independent of $r$. Explicitly, integrating against an arbitrary test function,
\begin{eqnarray}
\int_{-\infty}^{\infty}\lim_{r\to \infty}\frac{r^3}{(9x^2+1)^{\frac{d_s+6}{6}}}g(\tau)d\tau &=&2 \int_0^{\infty}\lim_{r\to \infty}\frac{1}{(9x^2+1)^{\frac{d_s+6}{6}}}g(\frac{x}{r^3})dx \\&=&
\frac{2\sqrt{\pi}\Gamma(\frac{3+d_s}{6})}{d_s\Gamma(\frac{d_s}{6})}g(0).
\end{eqnarray}

We therefore get a remarkably simple result for the Lorentzian Green's function defined by analytic continuation,
\begin{equation}
\phi = \frac{\phi_0 r^3}{(9 r^6 t^2 - 1)^{1+\frac{d_s}{6}}}. 
\end{equation}
Note that this has a singularity along $t = \pm \frac{1}{3 r^3}$, which corresponds to the radial null geodesics emanating from the point $t=0$ on the boundary; these are the light-cone singularities that we expect to see in the Lorentzian Green's function. To study the behaviour as $r \to 0$, we write 
\begin{equation} \label{xhor}
x= r^3t=r^3T+\frac{1}{3}+\frac{1}{2}R^2r^2
\end{equation}
so we have
 \begin{equation}
\lim_{r \to 0} \phi \sim \frac{r^{1-\frac{d_s}{3}}}{(6Tr+3R^2)^{1+\frac{d_s}{6}}} .
\end{equation}
This implies the solution becomes singular at the horizon for large spatial dimension $d_s$. 

Mathematically, the singularity at the horizon is related to the light-cone singularity:  the function $f(x)$ must have a singularity at $x = \pm \frac{1}{3}$, as this is the bulk light cone, but by  \eqref{xhor} we see that $x = \pm \frac{1}{3}$ on the (future/past) horizon as well, so the solution will also be singular there. There is an additional factor of $r^3$ in $\phi$ which vanishes on the horizon, but this is not sufficient to kill the singularity for large enough $d_s$. This mathematical relation makes it easy to see why we might expect the Green's function not to be regular on the horizon, but it's important to note that it's a mathematical relation, not a physical one; not all of the horizon is causally connected to the source, as discussed in the previous section. The singularity on the horizon is a new physical singularity, and not just the already noted light cone one.

The cases $d_s \leq 3$ seem special, as $\phi$ is then regular on the horizon for $R \neq 0$, although there is still a divergence as we approach the horizon for $R=0$. For $d_s=2$, which is physically the most interesting case, the Green's function on the horizon is proportional to $\delta(R)$:
 \begin{equation}
\lim_{r \to 0} \phi \propto T^{-1/3} \delta(R^2). 
\end{equation}
However, the finiteness of $\phi$ in these cases is somewhat misleading; if we consider the stress-energy tensor, we find that we can still expect a strong back-reaction on the metric. For a massless field,
\begin{equation}
T_{\mu\nu} = \partial_\mu \phi \partial_\nu \phi - \frac{1}{2} g_{\mu\nu} (\partial \phi)^2,
\end{equation}
and we find that in the new coordinates $T_{rr} \sim r^{-2d_s/3}$ as $r \to 0$ even for $R \neq 0$. Thus, there is a real singularity associated with this Green's function on the horizon.

Considering a spatially uniform source can lead to divergences even in cases where generic finite-energy excitations are regular on the horizon, as we see in appendix \ref{schrg}  for the  Schr\"odinger case. We therefore need to consider a spatially localized source. Unfortunately, this problem is more difficult, and we are not able to explicitly determine the Green's function. 

We consider again the massless Klein-Gordon equation, but now with a  boundary condition 
\begin{equation}
\lim_{r\to +\infty}\phi =C_1 \delta(t)\delta^{d_s}(\vec x)
\end{equation}
for some constant $C_1$. This boundary condition is covariant under the scaling symmetry satisfied by the equation \eqref{lscale}, so the solution should satisfy $\phi(\lambda t, \lambda \vec x, \lambda^{-1} r) = \lambda^{-(2d_s+3)} \phi(t, \vec x, r)$. Thus, the solution should be of the form $\phi = r^{2d_s+3} H(r^3 t, r^2 \rho)$, where $\rho^2 = x_i^2$ is a radial coordinate in the plane. Thus, finding the Green's function can be reduced to a problem in two variables,
\begin{eqnarray} \label{heq}
&& (9x^2-1)\frac{\partial^2H}{\partial x^2}+12xy\frac{\partial^2H}{\partial x \partial y}+(4y^2+1)\frac{\partial^2H}{\partial y^2}+ (15d_s+36)x\frac{\partial H}{\partial x}\\ \nonumber
&&+[(10d_s+22)y+\frac{d_s-1}{y}]\frac{\partial H}{\partial y}+ 3(d_s+2)(2d_s+3)H=0
\end{eqnarray}
where $\phi=r^{2d_s+3}H(x,y)$ and $x=r^3t$, $y=r^2\rho$. The form of this equation can be slightly simplified by a change of coordinates, 
\begin{equation}
\xi=\frac{x}{(1+4y^2)^{\frac{3}{4}}},\qquad \eta=y,
\end{equation}
which allows us to rewrite the equation as
\begin{eqnarray}
&& (9\xi^2-\frac{1}{\sqrt{1+4y^2}})\frac{\partial^2H}{\partial \xi^2}+(4y^2+1)^2\frac{\partial^2H}{\partial y^2}+ (9d_s+36)\xi\frac{\partial H}{\partial \xi}\\
&&+[(10d_s+22)y+\frac{d_s-1}{y}]\frac{\partial H}{\partial y}(4y^2+1)+ 3(d_s+2)(2d_s+3)(4y^2+1)H=0.\nonumber
\end{eqnarray} 
This transformation has eliminated the mixed derivative term. However, unlike in the Schr\"odinger case, this equation is still not separable, so we cannot solve for the Green's function exactly. 

We do have some general expectations for the singularity structure. Because of the non-relativistic causal structure of the boundary, the light-cone of a point on the boundary at $t=0$, $\vec x = 0$ is the same as the light cone of the surface $t=0$; thus we would expect that the Green's function will have singularities along the light cone $t = \pm \frac{1}{3r^3}$, that is at $x = \pm \frac{1}{3}$. The future horizon corresponds to $(x,y)\to (\frac{1}{3},0)$, so this light cone singularity leads us to expect that $H$ diverges on the horizon as well. Let us therefore assume that near the horizon, we have a leading singularity $H \sim (3x-1)^\alpha$. That is, assume a double Taylor expansion around $(x,y)\to (\frac{1}{3},0)$ of the form 
\begin{equation}
H = a (3x-1)^\alpha (1 + c_1 y + c_2 y^2 + c_3 (3 x-1) + c_4 y (3 x-1) + \ldots ).
\end{equation}
Noting that near the horizon $y = rR \sim \mathcal{O}(r)$, while $3 x -1 = \frac{3}{2} r^2 R^2 + 3 r^3 T \sim \mathcal{O}(r^2)$, the leading divergent terms in \eqref{heq} near the horizon are the $\frac{\partial^2 H}{\partial x^2}$ and $\frac{\partial H}{\partial x}$ terms, which go like $(3x-1)^{\alpha-1}$. This then fixes $\alpha$:
\begin{equation}
18 \alpha (\alpha -1) + 3 (5 d_s  + 12) \alpha = 0,
\end{equation} 
so $\alpha =0$ or $\alpha = -1 - \frac{5 d_s}{6}$. Taking the divergent solution, we would have 
\begin{equation}
\lim_{r \to 0} \phi \sim \frac{r^{2d_s +3}}{(3x-1)^{1+\frac{5 d_s}{6}}} \sim \frac{ r^{\frac{d_s}{3} +1}}{(3R^2 +6rT)^{1+\frac{5 d_s}{6}}}. 
\end{equation}
So the solution would be regular at $r=0$ and the stress tensor would be regular at $r=0$ for $R \neq 0$ for any $d_s$, with increasing numbers of derivatives regular as we consider larger dimensions. The difference from the spatially uniform case can be understood physically as the result of the spreading of the energy of the disturbance in the spatial directions. The solution has a singularity at $R=0$, where $\phi \sim r^{-d_s/2}$. In fact, this singularity is a delta function, as in the spatially uniform case for $d_s=2$: 
\begin{eqnarray}
\lim_{r\to 0} \phi &\sim& \lim_{r \to 0} \frac{ r^{\frac{d_s}{3} +1}}{(3R^2 +6rT)^{1+\frac{5 d_s}{6}}} 
\\ &\sim& \lim_{r \to 0} r^{\frac{d_s}{3} +1} \int_{0}^{+\infty}e^{-s(R^2+2rT)}s^{\frac{5 d_s}{6}}ds \\
&\sim& \lim_{r \to 0} r^{-\frac{d_s}{2}}  \int_{0}^{+\infty} e^{-\frac{wR^2}{r}-2wT}w^{\frac{5 d_s}{6}}dw 
\\ &\sim& \delta(R^2) T^{-\frac{d_s}{3}-1}
\end{eqnarray}
 Since the light-cone only intersects $r=0, R=0$ at $T=0$, this is not just the light cone singularity; there is a new physical singularity here. Since the solution is singular only at $R=0$, we would not expect the back-reaction of this scalar field divergence will be sufficient to obstruct an extension of the spacetime through the horizon for $R \neq 0$, but it may mean that the asymptotic behaviour at the whole of the other boundary cannot be maintained. If so, we would take it to mean that we can't physically calculate correlation functions between the two boundaries, so that the extension may not have a holographic interpretation in terms of full field theories living on the two boundaries. 
 
Clearly we have not established this divergence with any real rigour, and it would be useful to explore the behaviour of excitations in more detail. However, for the present Green's function analysis it is not clear that numerical solution of \eqref{heq} will be particularly useful, as the Green's function is really defined by satisfying the boundary condition in the Euclidean space and then analytically continuing to the Lorentzian section to evaluate it at the horizon. The best route to further work may be to look numerically for values of $\alpha$ such that the solution of the wave equation \eqref{generalequation} is regular in the interior and normalizable at infinity on both boundaries. We conjecture that no such values exist. 

\section*{Acknowledgements} This work is supported in part by STFC. We are grateful for useful conversations with Monica Guica, Jelle Hartong, Shamit Kachru, David Tong, Benson Way, and especially Sean Hartnoll, who pointed out \cite{Edgar} to us. 

\appendix

\section{Schr\"odinger Green's functions}
\label{schrg}

In section \ref{schr}, we studied the smoothness of extensions above the Schr\"odinger spacetimes by studying mode solutions. Another approach to considering whether the extension remains smooth for finite excitations is to consider instead an excitation created by acting with some localized source on the boundary. That is, we can ask if the boundary to bulk Green's function is smooth at the horizon. It is interesting to do this analysis for Schr\"odinger because in our analysis of  the hyperscaling violating case we work with this Green's function approach, so it is useful to have the corresponding results for Schr\"odinger for comparison.

One can give a simple abstract argument to suggest that the Green's function will remain smooth at the horizon in the case $z=2$; the geometry in the new coordinates \eqref{blaum} has a translation invariance in the $T$ direction, so the horizon at $T = \pi/2$ is not a special surface; if the Green's function insertion is at some arbitrary time, there is nothing to pick out this surface so the Green's function can't blow up there. 

However, this argument misses a subtlety, so it is useful to carry out an explicit analysis. We consider for simplicity the Green's functions for a massless scalar, $\nabla^2 \phi =0$. In the Schr\"odinger geometry with $z=2$, this is 
\begin{equation}
-\frac{2}{r^2} \partial_t \partial_\xi \phi + \partial_\xi^2 \phi + \frac{1}{r^2} \partial_{\vec{x}}^2 \phi + \frac{1}{r^{d_s+1}} \partial_r( r^{d_s+3} \partial_r \phi) = 0. 
\end{equation}
We will always assume that the solutions are plane waves in the $\xi$ direction, $e^{-im \xi}$, corresponding to considering sources carrying particle number proportional to $m$.  Consider first a source which is only localized in the time direction, and smeared uniformly with respect to $\vec{x}$. Then $\phi = e^{-im \xi} \phi(t,r)$, and 
\begin{equation} \label{seq1}
\frac{2im}{r^2} \partial_t \phi -m^2 \phi + \frac{1}{r^{d_s+1}} \partial_r( r^{d_s+3} \partial_r \phi) = 0. 
\end{equation}
We want to impose a boundary condition
\begin{equation}
\lim_{r \to \infty} \phi = e^{-im\xi} \delta(t),
\end{equation}
but we cannot impose such a delta-function boundary condition literally in the Lorentzian spacetime; the Lorentzian Green's function is divergent on the boundary not just at the point where the source is inserted but also at light like separation. The Schr\"odinger spacetime does not have an analytic continuation to a real Euclidean spacetime, but to model our hyperscaling violating calculation and to eliminate the $i$ in \eqref{seq1} it is convenient  to continue $t \to -i \tau$, so the wave equation becomes
\begin{equation}
-\frac{2m}{r^2} \partial_\tau \phi -m^2 \phi + \frac{1}{r^{d_s+1}} \partial_r( r^{d_s+3} \partial_r \phi) = 0.
\end{equation}
The key simplification that makes it possible to solve this equation in closed form is that the scaling symmetry under $t \to \lambda^2 t$, $r \to \lambda^{-1} r$ implies that the solution is of the form 
\begin{equation} \label{sff}
\phi = e^{-i m \xi} r^2 f(r^2 \tau). 
\end{equation}
Thus, the problem reduces to an ODE. Writing $x = r^2 \tau$, the equation for $f(x)$ is
\begin{equation}
4 x^2 \partial_x^2 f + (2 (d_s + 8) x - 2 m) \partial_x f + (2 d_s + 8 - m^2) f = 0. 
 \end{equation}
 The general solution is 
\begin{equation} \label{unisoln}
f(x) = c_1 x^{-\frac{d_s + 6}{4} -\frac{\nu}{2}} {}_1 F_1 (\frac{d_s + 6}{4} + \frac{\nu}{2}, 1 + \nu, -\frac{m}{2x} ) + c_2  x^{-\frac{d_s + 6}{4} +\frac{\nu}{2}} {}_1 F_1 (\frac{d_s + 6}{4} - \frac{\nu}{2}, 1 - \nu, -\frac{m}{2x} ),
\end{equation}
where $\nu^2 = \frac{ (d_s + 2)^2}{4} + m^2$. In the asymptotic region $r \to \infty$, the first term is the normalizable solution, and the second term is the non-normalizable solution. 

Since we want to impose a delta-function boundary condition, we want $\phi \to 0$ as $r \to \infty$ for $t \neq 0$, that is we want $\phi \to 0$ as $x \to \infty$, so we set $c_2 = 0$. The solution then approaches a distribution at the boundary $r \to \infty$. However, the solution is badly behaved at $\tau =0$ because of the divergence of \eqref{unisoln} when $x$ approaches zero from below, which prevents us from showing that the resulting distribution is in fact a delta function. This singularity at $t=0$ can be understood as the expected light cone singularity in the Lorentzian spacetime, since surfaces of constant $t$ are null surfaces in the Schr\"odinger spacetime. 
Analytically continuing back, the proposed Green's function is 
\begin{equation}
\phi = c e^{-im \xi} r^2  (r^2 t)^{-\frac{d_s + 6}{4} -\frac{\nu}{2}} {}_1 F_1 (\frac{d_s + 6}{4} + \frac{\nu}{2}, 1 + \nu, \frac{im}{2r^2 t} ).
\end{equation}
It is easy to see that this Green's function is also singular at the horizon $r \to 0$. The argument is the same as for the mode function in section \ref{schr}: the dependence on $\xi$ cannot be converted into dependence on the regular coordinate $V$.\footnote{In addition, using the asymptotic form of the confluent hypergeometric function given below, we would find that $\phi \sim r^{-d_s/2}$. } This is surprising in light of the previous abstract argument. The resolution is that we chose to put the source at $t=0$, which is a special point with respect to the horizon at $T = \pi/2$, and while the form of the source is invariant under the $t$-translation symmetry, it is not invariant under the $T$-translation symmetry, as this will act non-trivially on the $e^{-im\xi}$ factor. 

Physically, this divergence in the response to a spatially uniform source may be interpreted as the result of the harmonic potential in the $\vec{X}$ directions in the metric \eqref{blaum}. After half a period, this will cause particles starting at arbitrary values of $\vec{X}$ to become concentrated at a single point. 

Remarkably, for Schr\"odinger with $z=2$, we can go beyond this analysis for a spatially uniform source and explicitly construct the Green's function for a fully localized source. For a fully localized source, the scaling symmetry implies that the solution will be of the form $\phi = e^{-i m \xi} r^{2+d_s} f(r^2t, r \vec{x})$. As before, we make an analytic continuation to set $t = -i \tau$, and write the solution as 
\begin{equation}
\phi = e^{-i m \xi}r^{2+d_s} f(x,y),
\end{equation}
where $x = r^2 \tau$ as before and $y = |\vec{x}|^2 /\tau$. The equation of motion then becomes
\begin{equation}
4 x^2 \partial_x^2 f + (2 (3d_s + 8) x - 2 m) \partial_x f + (2 d_s^2 + 8 d_s + 8 - m^2) f  + \frac{1}{x} \left( 4y \partial_y^2 f + (2 d_s +2m y) \partial_y f \right) =0 .
\end{equation}
This is separable. What is more, there is a separable solution which approaches a distribution supported just at $\tau =0$, $\vec{x} = 0$. This solution is 
\begin{equation}
\phi = c e^{-im\xi - \frac{m y}{2}} r^{2+d_s} x^{  -\frac{3}{4}(d_s + 2) - \frac{\nu}{2}} {}_1 F_1(\frac{d_s +6}{4} + \frac{\nu}{2}, 1 + \nu,-\frac{m}{2x} ), 
\end{equation}
where $\nu$ is as before. To see that this satisfies the boundary conditions, note that the $x$ dependence makes it vanish as $r \to \infty$ for $\tau \neq 0$ as before, so $\phi$ is supported only at $\tau=0$ in the limit, and then that as $\tau \to 0$, $e^{-my/2} \to 0$ for $\vec x \neq \vec 0$, so $\phi$ is supported only at $\vec{x} = 0$ in the limit. 

Thus, analytically continuing back in $t$, the candidate Lorentzian Green's function is 
\begin{equation}
\phi = c e^{-im\xi +i \frac{m \vec x^2 }{2 t}} r^{2+d_s} (r^2 t)^{  -\frac{3}{4}(d_s + 2) - \frac{\nu}{2}} {}_1 F_1(\frac{d_s +6}{4} + \frac{\nu}{2}, 1 + \nu,\frac{im}{2 r^2 t} ).
\end{equation}
Again, this is singular at $t=0$, which is the light-cone singularity in spacetime. This is singular at $t=0$ for all $\vec x$ even for a source which is localized at $\vec x=0$ because of the non-relativistic causal structure: all points at $t=0$ are lightlike separated from this boundary point. To examine the behaviour near the horizon at $r \to 0$, use the asymptotic expansion of the confluent hypergeometric function \cite{Temme}
 \begin{equation}
_1F_1(a,b,z)\sim \frac{\Gamma(b)}{\Gamma(a)}e^zz^{a-b}+\frac{\Gamma(b)}{\Gamma(b-a)}(-z)^{-a}
\end{equation}
which gives 
\begin{equation}
\phi = e^{-im\xi +i \frac{m \vec x^2 }{2 t}}  \left( c' (rt)^{-d_s-2} e^{\frac{im}{2 r^2 t}} + c'' r^{\frac{3}{2} d_s + 2} (rt)^{-\frac{d_s}{2}} \right).
\end{equation}
Making the coordinate transformation \eqref{bcoord}, we see $r t = \frac{\sin T}{R}$ is finite as $T \to \pi/2$, so the second term vanishes and the first term is finite. Using \eqref{bc2}, the combination appearing in the exponential is 
\begin{equation}
\xi - \frac{\vec x^2}{2t} - \frac{1}{2 r^2 t}  = V -  \frac{1}{2} (R^2 + \vec{X}^2) \cot T
\end{equation}
so as $T \to \pi/2$, 
\begin{equation}
\phi \approx c' e^{-imV} R^{d_s+2}
\end{equation}
is perfectly regular. 

This Green's function analysis does not extend simply to $z >2$. The particle number $m$ has a non-zero scaling dimension, so the previous argument that $\phi$ will only involve a function of $r^z t$ and $r \vec x$ does not apply; the function can depend separately on $r, t, \vec{x}$ with the scaling being soaked up by appropriate powers of $m$. Thus even the spatially uniform source will involve solving a PDE, and we have not explored the problem further.

\section{Green's function for $n>2$}
\label{gen}

In this appendix, we consider the Green's function analysis for general $n$. The results are essentially the same as for $n=2$. For a spatially uniform boundary condition, the scaling argument determines $\phi = r^{2n-1} f(r^{2n-1} t)$ and the massless scalar field equation reduces to the ODE \begin{equation}
[(2n-1)^2x^2-1]f''(x)+(2n-1)(8n-4+d_s)xf'(x)+(2n-1)(d_s+4n-2)f(x)=0,
\end{equation}
where $x=r^{2n-1}t$.  As we are approaching the horizon, we can use (\ref{nTcoord}) to rewrite this in terms of the regular coordinates, \begin{equation}
x=r^{2n-1}t = r^{2n-1}T+\frac{1}{2n-1}+\frac{r^{2n-2}R^2}{2}.
\end{equation}
The solution of the equation again has a simple form:
\begin{equation}
f(x)=[(2n-1)^2x^2-1]^{-1-\frac{d_s}{2(2n-1)}}.
\end{equation}
This means \begin{eqnarray}
\lim_{r\to 0} \phi &=& \frac{\phi_0 r^{2n-1}}{[(2n-1)^2x^2-1]^{1+\frac{d_s}{2(2n-1)}}} \\
&\sim & \frac{r^{1-\frac{d_s(n-1)}{2n-1}}}{(R^2+2rT)^{1+\frac{d_s}{2(2n-1)}}},
\end{eqnarray}
which is generally divergent at $r \to 0$ for high enough $d_s$. The effect of increasing $n$ is to strengthen the divergence.

For a fully localized source, the solution should be $\phi=r^{2n-1+d_sn}H(x,y)$, and we will have the equation \begin{eqnarray}
&& [(2n-1)^2x^2-1]\frac{\partial^2 H}{\partial x^2} +2n(2n-1)xy \frac{\partial^2 H}{\partial x \partial y}+  (n^2y^2+1)\frac{\partial^2 H}{\partial y^2} \\ \nonumber 
&+& (2n-1)(8n-4+2d_sn+d_s)x \frac{\partial H}{\partial x} + [(7n^2-3n+(2n^2+n)d_s)y+\frac{d_s-1}{y}] \frac{\partial H}{\partial y} \\&+& (2n-1+d_s n)(4n+d_s+nd_s-2)H=0 \nonumber 
\end{eqnarray}
Making the assumption that $H$ has a divergence proportional to $[(2n-1) x -1]^a$ as $r \to 0$, it will again be the $\frac{\partial^2 H}{\partial x^2}$ and $\frac{\partial H}{\partial x}$ terms that dominate, determining $a$. 
The solution would then behave as  \begin{eqnarray}
\lim_{r_\to 0} \phi  &\sim& \frac{r^{2n-1+d_s n}}{[(2n-1)x-1]^{1+\frac{(2n+1)d_s}{2(2n-1)}}} \\
&\sim& \frac{r^{1+\frac{d_s}{2n-1}}}{(R^2+2rT)^{1+\frac{(2n+1)d_s}{2(2n-1)}}}.
\end{eqnarray}
This again gives a delta-function singularity at $r \to 0$, $\lim_{r \to 0} \phi \sim \delta(R^2) T^{-1-\frac{d_s}{2n-1}}$.

\end{document}